\documentclass[5p,10pt,sort&compress]{elsarticle}
\usepackage[USenglish]{babel}
\usepackage{enumerate}
\usepackage{nth}
\usepackage{mhchem}
\usepackage{amsmath}  
\usepackage{amssymb}  

\usepackage{mathtools}
\usepackage{siunitx}

\DeclareMathOperator\erf{erf}
\usepackage{graphicx}  
\usepackage{verbatim}  
\usepackage{color}   
\usepackage{subfigure} 
\usepackage{hyperref}  

\usepackage{listings}

\usepackage{amssymb}

\newcounter{bla}

\journal{Computer Physics Communications}

\begin{document}
\begin{frontmatter}



\title{FLAME: a library of atomistic modeling environments}


\author[a]{Maximilian Amsler\corref{author}}
\author[b]{Samare Rostami}
\author[b]{Hossein Tahmasbi}
\author[b]{Ehsan Rahmatizad}
\author[b]{Somayeh Faraji}
\author[b]{Robabe Rasoulkhani}
\author[b]{S. Alireza Ghasemi\corref{author}}

\cortext[author] {Corresponding authors.\\\textit{E-mail address:} amsler.max@gmail.com, aghasemi@iasbs.ac.ir}
\address[a]{Laboratory of Atomic and Solid State Physics,
Cornell University, Ithaca, New York 14853, USA}
\address[b]{Department of Physics, Institute for Advanced Studies in Basic Sciences (IASBS), Zanjan 45137-66731, Iran}

\begin{abstract}
FLAME is a software package to perform a wide range of atomistic simulations
for exploring the potential energy surfaces (PES) of complex condensed matter systems.
The range of methods include molecular dynamics simulations to sample free energy landscapes, 
saddle point searches to identify transition states, and gradient relaxations
to find dynamically stable geometries.
In addition to such common tasks, FLAME implements a structure prediction algorithm
based on the minima hopping method (MHM) to identify the ground state
structure of any system given solely the chemical composition, and a
framework to train a neural network potential to
reproduce the PES from \textit{ab initio} calculations.
The combination of neural network potentials
with the MHM in FLAME allows a highly
efficient and reliable identification of the ground state
as well as metastable structures  of molecules and crystals, 
as well as of nano structures, including surfaces, interfaces, 
and two-dimensional materials.
In this manuscript, we provide detailed descriptions
of the methods implemented in the
FLAME code and its capabilities, together with
several illustrative examples. 
\end{abstract}


\end{frontmatter}


\noindent
{\bf PROGRAM SUMMARY/NEW VERSION PROGRAM SUMMARY}

\noindent
{\em Program Title:} \texttt{FLAME}                                          \\
{\em Licensing provisions:} GPLv3                                   \\
{\em Programming language:} Fortran90, Python                                   \\
{\em External routines/libraries:} \texttt{BigDFT PSolver}~\cite{Genovese2008,Genovese2006,Neelov2007,Genovese2007}, \texttt{Spglib}~\cite{togo_spglib_2018}, MPI, LaPack, Blas\\
{\em Program summary URL:} \url{http://flame-code.org}\\
{\em Program obtainable from:}  \url{https://github.com/flame-code/FLAME}\\
{\em Computer:} The program should work on any system with a F90 compiler. The code has been tested with the Intel and GNU Fortran compilers.\\
{\em Operating system:} Unix/Linux\\
{\em RAM:} several GB\\
{\em No. of lines in distributed program, including test data, etc.:} 231 633\\
{\em Nature of problem:} Exploring the potential energy landscapes of complex condensed matter systems, their stationary points, and their global minima.\\
{\em Solution method:} A neural network representation of the potential energy landscape in conjunction with a library of methods to explore its features, most notably the minima hopping approach.\\
   \\

\section{Introduction}
A wealth of materials properties is intrinsically
encoded in the topology of the (free) energy
landscape of a condensed matter system.
In solids, kinetically persistent atomic structures 
correspond to local minima on the potential energy
surface (PES), while the global minimum on the PES
represents the ground state structure.
In chemical reactions or phase transitions,
the reaction coordinate follows the minimal
energy pathway (MEP) between two states (local minima)
and passes through a first order saddle point.
Identifying these stationary points on the PES is of importance in physics,
chemistry, and materials science, since the structure
of matter fundamentally governs the physical/chemical properties of materials, 
including crystal phase stability, solubility, optical and transport phenomena,
elastic properties, and many more.

Characterizing relevant portions of a PES through atomistic simulations poses
two non-trivial tasks, namely (a) finding an accurate representation of the PES and
(b) sampling it efficiently:

\subsection*{(a) Representation}
 The task of representing the PES requires the
mapping of the atomic positions (and the cell vectors for periodic systems)
to a scalar total energy based on the underlying atomic interactions.
The most accurate methods to compute this energy involve solving
in some manner the electronic Schroedinger equation
within the Born-Oppenheimer approximation.
Over the last decades, 
density functional theory (DFT) has
evolved as a leading method to tackle this challenge
due to its convenient accuracy
at a relatively moderate computational effort.
Nevertheless, the current computer resources
limit its application to at most hundreds of atoms
when the PES has to be densely sampled through many
evaluations, e.g., in structural searches or for molecular
dynamics (MD) simulations.

To model larger systems or to
accelerate atomistic simulations, alternative methods based on approximate,
classical or semi-classical approaches are required,
e.g., force fields and tight-binding schemes.
In the classical approaches, atomic interactions
are approximated by analytical functions justified by the types of
the underlying chemical bonds~\cite{Stillinger1985,Tersoff1989,Justo1998,Lenosky2000}.
Commonly, these interactions include attractive short-range contributions, 
like stretching, bending, and torsional terms, and non-bonding electrostatic 
or van der Waals interactions, together with some 
repulsive terms. 
Such potentials have to be parameterized and fitted to either
experimental or to accurate \textit{ab initio} data, which
limits their application to systems with predetermined compositions 
and a small set of distinct elements.
Nevertheless, modern empirical potentials that include
charge transfer algorithms have been applied to a wide range of
materials~\cite{vanDuin2001,Yu2007,Liang2013}.

Recently, with the advent of machine learning (ML) algorithms,
there has been a surge in interest for using ML
techniques to interpolate the first principles
PES by training them on large
sets of reference data~\cite{Behler2007,Bartok2010,Rupp2012,Thompson2015,Ghasemi2015,gap}.
Due to their flexible functional forms, such ML models
are capable to reproduce the \textit{ab initio} results
with very high accuracy. However,
regions of the configurational space that
are not well sampled in the training data set
may be poorly described
given that ML interatomic potentials are (sophisticated)
regression models.
Hence, the non-physical form of ML potentials
is both a blessing (for interpolation) and a curse (for extrapolation).
\subsection*{(b) Sampling} Sampling the PES is challenging due to the
high dimensionality $D$ of the PES, which is a function
of the atomic (and cell) degrees of freedom.
For a molecular system with $3N_{at}$ atoms, $D$ corresponds
to $3N_{at}-6$ (the Cartesian coordinates
$\mathbf{r}_1,\mathbf{r}_2,\dots,\mathbf{r}_{N_{at}}$
of the atoms, taking into account rotational and
translational invariance), while for a crystalline/periodic
system $D=3N_{at}+3$ (which includes the reduced atomic
coordinates $\mathbf{s}_1,\mathbf{s}_2,\dots,\mathbf{s}_{N_{at}}$ with
$\mathbf{r}_i = h\mathbf{s}_i$
and $9$ components of the cell vectors $h=\{\mathbf{a},\mathbf{b},\mathbf{c}\}$,
subject to rotational and translational invariance).
Further, the complexity of the PES scales dramatically
with the number of degrees of freedom, e.g.,
the number of local minima on a PES increases
exponentially with system size~\cite{stillinger_exponential_1999}, which makes
structure prediction challenging for
any realistic system~\cite{oganov_modern_2010}.

A plethora of packages have been
developed in recent years to tackle above tasks separately. 
AMP~\cite{amp}, PES-Learn~\cite{peslearn}, 
and TensorMol~\cite{tensormol} are only a few examples of 
codes to generate and train ML models
of PESs. They offer optimized workflows to
extract structural features and attribute which are used 
as inputs to train artificial neural network (ANN) or Gaussian process models to accurately reproduce 
\textit{ab initio} results. Generally, these codes have
to be subsequently coupled to
external packages which implement advanced sampling algorithms,
like LAMMPS~\cite{plimpton_fast_1995} or ASE~\cite{Hjorth_Larsen_2017}.
Other  codes specialized on PES sampling alone
with a particular aim at structure prediction  
range from those based on genetic algorithms 
(USPEX~\cite{glass2006uspex}, XtalOpt~\cite{lonie2011xtalopt})
and particle swarm optimization (Calypso~\cite{wang2012calypso}) to
random searches AIRSS~\cite{pickard2011ab}.

FLAME provides the tools to tackle both challenges
within a fully integrated package.
In terms of sampling schemes,
we have implemented molecular dynamics, saddle point search methods,
and, most notably, an efficient structure prediction method based
on the MHM.
These methods can be coupled to any scheme to evaluate the PES, and we have 
incorporated interfaces for a range of DFT packages and molecular mechanics tools.
At the same time, we have implemented a highly efficient and
accurate ANN potential to
approximate a PES by training it to \textit{ab initio} reference data.
This synergy of an efficient structure prediction method together with 
an ANN potential is a particularly powerful feature of FLAME,
which significantly accelerates global geometry optimizations and
allows the study of larger, increasingly complex and  realistic systems.

Beside the ANN as the central technique, FLAME brings along a range of integrated
interatomic potentials, including
the environment dependent interactive potential (EDIP)~\cite{Justo1998},
Lenosky~\cite{Lenosky2000} and Tersoff~\cite{Tersoff1989} potentials for silicon.
The sampling methods in FLAME can be linked with external codes through 
sockets, allowing communication across the internet or local UNIX-domain sockets. 
For this purpose, the i-Pi protocol is implemented, where FLAME acts as a server
and the software packages evaluating the PES
act as clients~\cite{ceriotti_i-pi:_2014}.
Conversely, FLAME can act as a client as well,
conveniently through the i-Pi socket protocol
In this way, 
the methods to evaluate the PES 
within FLAME can be seamlessly integrated into external codes that
provide their own sampling algorithms, e.g., 
LAMMPS~\cite{plimpton_fast_1995}.

This manuscript describes the key features of FLAME and is structured as follows:
Sec.~\ref{sec:io} briefly describes the input and 
output file formats of FLAME, Sec.~\ref{sec:ann}
outlines the implementation of
the ANN schemes in FLAME, while the relevant
sampling algorithms are described in Sec.~\ref{sec:PES}.
Each section is accompanied with detailed examples and
suggested input parameters, if applicable.
In Sec.~\ref{sec:descriptors}
we review atomic environment descriptors and
configurational fingerprints which are
implemented in FLAME.
We present a brief description of the implemented methods
to efficiently calculate the Hartree energy in Sec.~\ref{sec:electrostatics}.
Finally, we conclude with a summary and an outlook in Sec.~\ref{sec:conclusions}.

\section{Input and Output Files\label{sec:io}}
FLAME takes no command-line arguments but uses structured input and output files 
in case-sensitive YAML format for convenient scripting and post-processing.
The main input and output files are called
\texttt{flame\_in.yaml} and \texttt{flame\_log.yaml}, respectively,
and contain hierarchical keyword--parameter pairs.
The \texttt{main} block in \texttt{flame\_in.yaml}  
embeds
the most important input key, \texttt{task},
which determines what kind of 
atomistic simulation to run within the 
FLAME executable. Further, this block also
includes general parameters that describe the system
and the computational setup,
such as the atomic types, applied pressure, the output verbosity, etc. 
Essentially, the \texttt{main} block sets up the atomistic 
modeling environment.

The \texttt{potential} block determines the method
used to model the atomic interactions. The associated 
keyword \texttt{potential} can range from a name of 
an integrated force field (e.g.,  \texttt{lj} for the Lennard-Jones potential)
to the name of an externals software package (e.g., \texttt{dftb} for
the DFTB+ package~\cite{aradi_dftb+_2007}, or \texttt{netsock} for network sockets). Additional parameters related to 
the potential are included here, such as k-points density, cutoff radii,
and how to treat long-range electrostatic interactions.

Additional blocks and subblocks can be added to specify the detailed parameters
of the simulation. E.g., the method determined in the keyword \texttt{task} 
itself defines a
subblock with its own subparameters. All possible blocks, their keywords, and the 
associated parameters are printed at the head of the output \texttt{flame\_log.yaml} file,
thereby allowing a rapid assessment of available
and relevant input options.
A detailed description of the
various keywords can be found in the manual.

The native format of FLAME to handle atomic structure files is based on YAML 
as well. The default input and output structure filenames are \texttt{posinp.yaml}
and \texttt{posout.yaml}, respectively. FLAME brings along a set of convenient python scripts to inter-convert between YAML and other common structure formats, including \texttt{XYZ}, \texttt{ascii}, \texttt{POSCAR}, and many more.

\subsection{Example}
A sample input file for a molecular dynamics simulation
with an \textit{NVT} ensemble of silicon is presented in Fig.~\ref{fig:code_md}.
The \texttt{main} block determines that a \texttt{dynamics}
simulation is to be performed with silicon \texttt{types} of atoms.
The employed atomic \texttt{potential} is the Lenosky tight-binding model~\cite{Lenosky2000}, 
\texttt{ltb}. \texttt{dynamics} itself spawns a block,
where the method \texttt{md\_method}, the time-step \texttt{dt} in fs, and the number of iterations
\texttt{nmd} are specified. 
We set the initial and target temperature to  $\texttt{init\_temp} =  \texttt{temp} = 300$~K 
using the Nose-Hoover (\texttt{nvt\_nose}) method  with $\texttt{ntherm} = 2$  chained thermostats.

\begin{figure}[htb]
\begin{lstlisting}[frame=single]
main:
    task:         dynamics
    types:        Si
potential:
    potential:    ltb
dynamics:
    md_method:    nvt_nose
    dt:           2.0
    nmd:          10000
    init_temp:    300.0
    temp:         300.0
    ntherm:       2
    highest_freq: 10
\end{lstlisting}
\caption{A sample input \texttt{flame\_in.yaml} file for an \textit{NVT} molecular dynamics simulation of a silicon system.~\label{fig:code_md}}
\end{figure}

Each of the following sections
describes in detail the core
functionalities of FLAME associated to the
\texttt{task} keyword.
\section{Neural Network Potentials\label{sec:ann}}
In recent years, a new generation of interatomic potentials
have been introduced that are based on ML techniques.
These potentials are not restricted by a predefined functional form,
but are composed of highly flexible multivariate functions with 
parameters that are optimized during
a fitting process, referred to as ``training''.
One type of such an ML potential approach is based on artificial neural networks (ANN)
that is inspired by the neural systems in living organisms and
imitates how information is passed between 
neurons through synapses.

In an ANN, the artificial neurons are represented by nodes, which are 
connected with each other to pass information, the artificial synapses, referred to as edges.
Commonly, the nodes are arranged in multiple layers,
each containing several nodes.
The input data is fed into the ANN through its ``input layer'' and, after being processed, passes
to the next layer, called the ``\nth{1} hidden layer''.
The output of this layer is then fed into
the next hidden layer, and so on, until the 
last hidden layer is reached, eventually arriving at the 
``output layer''. The training of an ANN involves fitting the 
parameters that determine the model by adjusting the weights of the ANN.

Early versions of ANN interatomic potentials 
used to directly feed the Cartesian coordinates of some or all atoms
in a system into the input layer,
contained only one node in the output layer,
and its value was trained on the total energy of
the particular atomic arrangement~\cite{Lorenz2004}.
Two major shortcoming of this approach were quickly 
recognized:
(i) the number of atoms had to remain unchanged for both the  
training and prediction tasks, hence the ANN 
wasn't transferable to different stoichiometric conditions,
(ii) and the Cartesian coordinates are 
ill suited as input parameters since they are not invariant under rotation
and translation of the structure, while such transformations
must preserve the total energy.

In 2007, Behler and Parrinello~\cite{Behler2007} (BP)
introduced a new approach to address above problems.
In their method, the total energy was
expressed as a sum of atomic energies.
Each atomic energy $E_i$ is thus obtained through an
ANN process which is fed with information of the
environment of the corresponding atom with index $i$,
thereby alleviating problem (i).
Problem (ii) was tackled by first
mapping the Cartesian coordinates to an array of
values describing the local environment of each atom, 
referred to as an atomic environment descriptor.
This transformation is indeed an essential component 
for any modern ML interatomic potential, and
its detailed procedure 
can significantly affect the performance.
Since every part of the BP approach is local, 
it is less suited for systems
in which long range interactions are
of importance, e.g.,
for ionic systems.
To improve accuracy, a Coulombic term
can be added to the total energy in order to account for the electrostatic
interactions~\cite{Artrith2011}.

In FLAME, one can select the BP approach by setting
\texttt{approach} in block \texttt{ann}
to \texttt{atombased}.
More details on environment descriptors can be found in Sec.~\ref{sec:descriptors}.

\subsection{Charge Equilibration via Neural Network Technique}
\begin{figure}[htb]
\centering
\includegraphics[width=1\columnwidth]{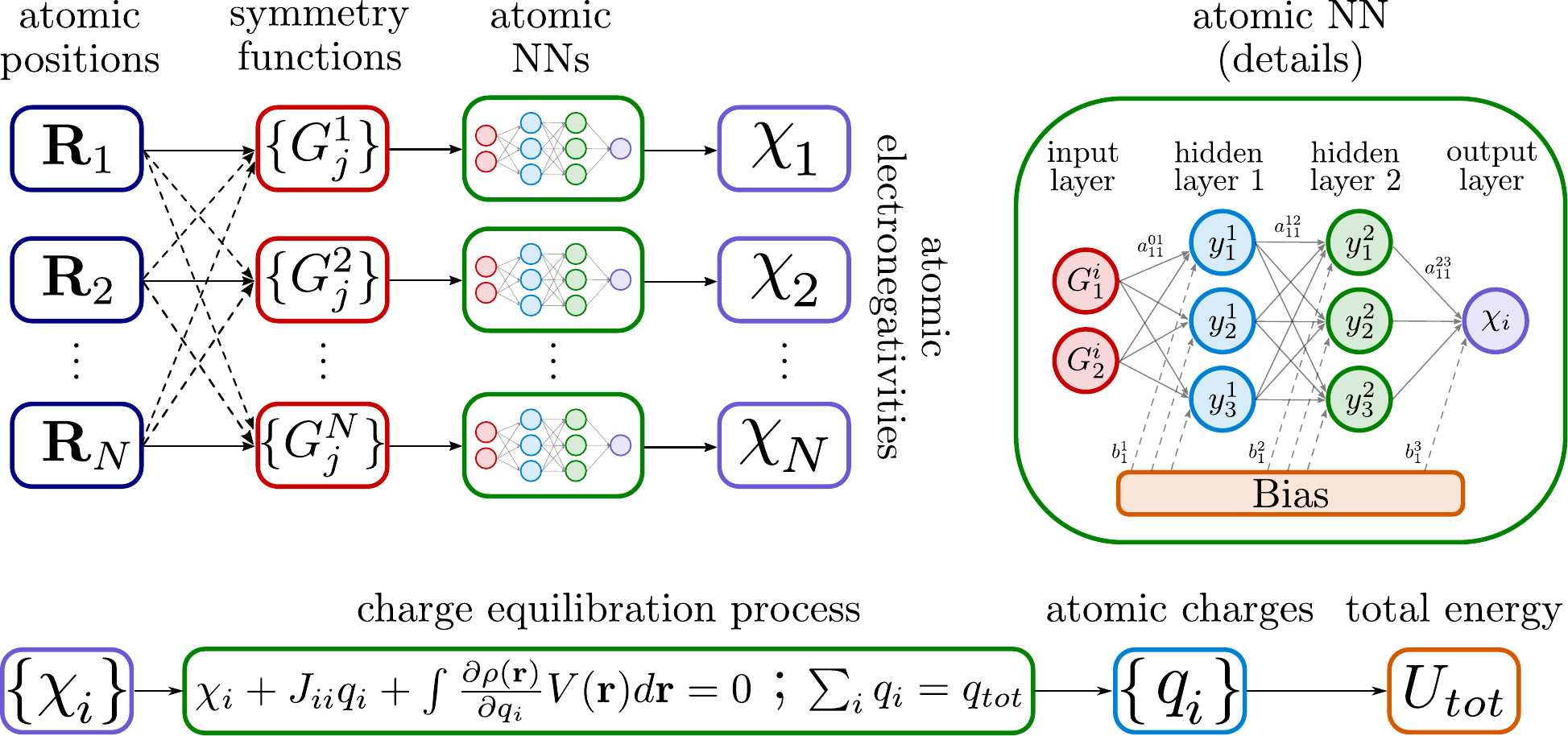}
\caption{Schematic illustration of the CENT method}\label{fig:cent}
\end{figure}
In 2015, Ghasemi~\textit{et al.} introduced an alternative approach of employing 
ML to model atomic interactions. Instead of directly predicting the energy of a system, an ANN process is applied for each atom 
to predict an environment-dependent atomic electronegativity in the output layer,
which in turn is fed into an energy functional~\cite{Ghasemi2015}.
This method was coined the ``charge equilibration via neural network
technique'' (CENT) and has been successfully applied to a variety of
predominantly ionic materials, including NaCl~\cite{Ghasemi2015},
\ce{CaF2}~\cite{Faraji2017,Faraji2019}, ZnO~\cite{Rasoulkhani2017},
\ce{TiO2}~\cite{Asna2017}, \ce{WS2}~\cite{Hafizi2017} and
six-component alkali halide compounds~\cite{Rostami2018}.
Overall, errors in physical and chemical
quantities investigated in these studies
lie well within
a few percent with respect to DFT reference values, mostly less than or comparable to the variations among
different exchange-correlation functionals employed in DFT calculations.
The components of the CENT method, including the high-dimensional
neural networks and the charge equilibration process, are depicted
in Fig.~\ref{fig:cent}.
Here, we give a brief review of the CENT method.

For a system consisting of $N$ atoms,
we express the total energy as
\begin{multline}
U_{tot}(\left\{q_{i} \right\})=
\sum_{i=1}^{N} \left(E_i^0 + \chi_{i} q_{i} + \frac{1}{2} J_{ii} q_i^2 \right)\\ +  \label{eqn:etot1}
\frac{1}{2} \iint \frac{\rho({\bf r}) \rho({\bf r}')}{|{\bf r}-{\bf r}'|}
\;d{\bf r}\;d{\bf r}',
\end{multline}
where $E_i^0$ is the energy of the individual, isolated atom $i$, and
$\chi_i$ is the environment dependent atomic electronegativity
of atom $i$ whose functional dependence is determined by an ANN.
$q_i$ and  $J_{ii}$ are the corresponding atomic charge and the element-dependent
atomic hardness~\cite{Mortier1985}, respectively, and
$\rho(\textbf{r})$ is the charge density of the system,
which, in our implementation, is given by a superposition of
spherical Gaussian functions centered at the atomic positions
$\textbf{r}_i$, each normalized to the corresponding atomic
charge $q_{i}$.
For non-periodic systems, Eq.~\eqref{eqn:etot1} can be rewritten
in a simple form:
\begin{multline*}
U_{tot}(\lbrace q_{i}\rbrace,\lbrace \textbf{r}_{i}\rbrace)=
\sum _{i=1}^{N}\left[ (E_{i}^{0} +\chi_{i} q_{i} +
\frac{1}{2}( J_{ii}+\frac{2\gamma_{ii}}{\sqrt{\pi}}) q_{i}^{2}\right]\\+
\sum _{i>j}^{N}q_{i}q_{j}\frac{\erf(\gamma_{ij}r_{ij})}{r_{ij}},
\end{multline*}
where $\gamma_{ij}=\frac{1}{\sqrt{\alpha_{i}^{2}+\alpha_{j}^{2}}}$ and
$\alpha_{i}$ are the widths of the Gaussian atomic charge densities,
and $r_{ij}$ is the distance between the atoms $i$ and $j$.
The energy functional must be minimized with respect to
the charge density which, in our scheme, is done by
minimizing $U_{tot}$ with respect to the $q_{i}$'s,
similar to
a charge equilibration process.
The minimization is performed under the constraint of fixed total
charge to a constant value using Lagrange multipliers, which
leads to a system of linear equations of the form
$\tilde{A}{\bf Q}=-{\bf \chi}$,
where $\tilde{A}$ is a $(N+1)\times (N+1)$ matrix, and
${\bf Q}$ and ${\bf \chi}$ are $(N+1)$-dimensional vectors.
For non-negative values of $J_{ii}$'s, it is guaranteed
that the matrix of our system of linear equations
is non-singular. With this approach, charge can transfer in a long range manner
while the total charge of the system is conserved.
Note that the atomic charges $q_{i}$
are implicitly environment dependent
through the atomic electronegativities $\chi_i$.

To solve the system of linear equations using iterative methods,
we need the gradient of the total energy with 
respect to the atomic charges.
Differentiating Eq.~\eqref{eqn:etot1} with respect to $q_i$
we obtain the gradient
\begin{align*}
g_i=\frac{\partial U_{tot}}{\partial q_i}=\chi_i + J_{ii}q_i + g_i^{(h)},
\;\;\forall\, i=1, \dots , N
\end{align*}
where $g_i^{(h)}$ is the contribution from the
Hartree energy and is given by
\begin{align*}
g_i^{(h)} = \int \frac{\partial \rho(\textbf{r})}{\partial q_i} d\textbf{r}
\int \frac{\rho(\textbf{r}')}{|\textbf{r}-\textbf{r}'|} d\textbf{r}'=\int
\frac{\partial \rho(\textbf{r})}{\partial q_i}
V(\textbf{r}) d\textbf{r}.
\end{align*}
The potential function, $V(\textbf{r})$, can be obtained by solving
the Poisson's equation with the appropriate boundary conditions (BC) of the problem.
A discussion of the electrostatic methods implemented in FLAME is given
in Sec.~\ref{sec:electrostatics}.

Once the gradient is computed, the system of linear equations
can be solved.
The constraint of fixed total charge is fulfilled by using the modified gradient given by
$g_i-\frac{1}{N}\sum_{l=1}^N g_l$.
In the case of free BC and a small number of
atoms the system of equations can be solved by means of direct methods,
whereas for large systems it is convenient to use an iterative scheme.
In contrast, for bulk structures the system of equations is always solved
iteratively, irrespective of the system size.
Since the system of linear equations is well-conditioned
the total number of iterations to reach sufficient convergence
rarely exceeds $100$ based on extensive tests for small and medium sized systems.
Also, it is possible to significantly reduce this number when performing
molecular dynamics simulations or local geometry relaxations
with relatively small atomic displacements in consecutive time step
since the initial guess for the atomic charges can be taken
from the converged values obtained in the previous step.

\subsection{Potential Training\label{sec:training}}
Since ANN ML models do not have a functional form and
contain many parameters one may very easily 
encounter issues due to over-fitting.
Therefore, in contrast to usual force fields, one must generate
a large number of reference data points ranging from
thousands to tens of thousands of configurations.
More precisely, the training data must be sufficiently diverse and extensive to prevent over-fitting.
In fact, the most challenging task when constructing
an accurate and transferable ML potential
is generating a suitable reference data set.

We commonly generate such a data set in several steps.
First, we start with a small set of configurations that is
generated using DFT calculations based on one of the following methods:
\begin{enumerate}[i]
\item  \textit{ab initio}  molecular dynamics simulations starting from
different well-known structures at the given composition.
\item  random structures that are relaxed to within a very loose
tolerance, i.e., by performing only a few iterations as well as using
loose input parameters of the \textit{ab initio} package.
\item  elemental substitution in structural prototypes (e.g., such obtain in earlier fitting data sets or online structure repositories)
together with an appropriate scaling of the interatomic distances based on the atomic radii.
\end{enumerate}

This small data set is then used to construct a first, approximate CENT potential
with limited accuracy.
In a next step, this preliminary potential is used in multiple 
structure prediction runs to sample the PES
with various system sizes and starting from different seed configurations.
For this purpose we employ the MHM, which not only tries to find the global minimum but also efficiently explores low-lying portions of the PES (see Sec.~\ref{sec:MH} for details).
In this way, a large number of new structures are generated that
can be used to extend the initial training data set.
Since the approximate potential trained to the first, small training set
can produce nonphysical structures, 
we have to exclude them and filter for structures with, e.g., unreasonable bond lengths.
We further screen the data set for
configurations that are too similar to each other
by using 
distances of atomic environment descriptors
or structural fingerprints
in order to retain a high structural diversity.
Finally, DFT calculations are performed on the new configurations to 
update the training data set and construct
a more accurate CENT potential.

\begin{figure}[htb]
\centering
\includegraphics[width=0.7\columnwidth]{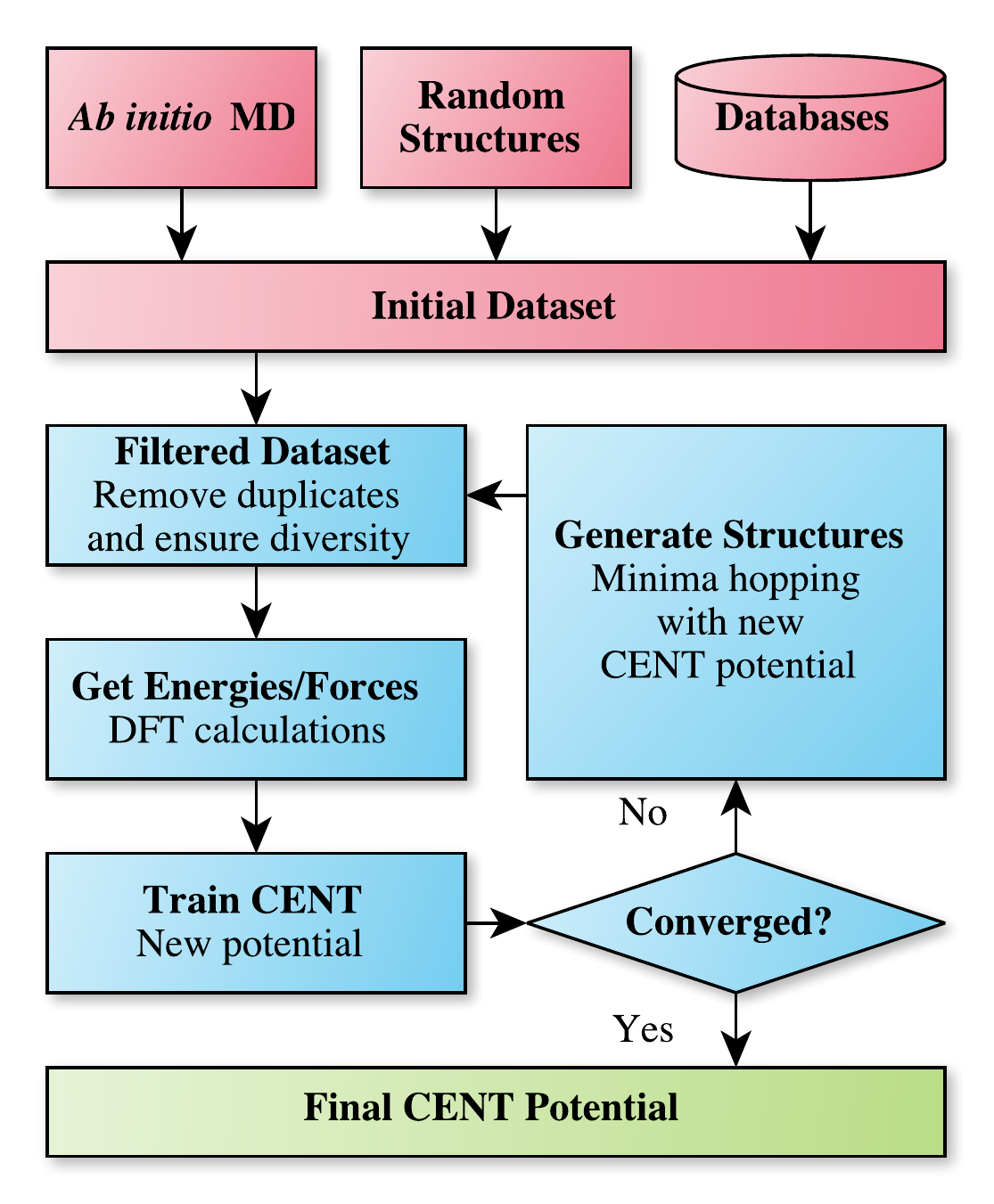}
\caption{Flowchart of the CENT training procedure.
\label{fig:training}}
\end{figure}

In practice, we repeat this process of refining the reference data several times until the training
set is sufficiently large and contains diverse structures to reach the desired  accuracy and reliability
of the resulting CENT potential.
All the steps involved in the procedure above are fully implemented in FLAME, i.e.,
training an ANN potential, excluding similar structures, performing MHM or MD runs to
generate new structures, etc.
Fig.~\ref{fig:training} shows a flowchart of the iterative CENT training algorithm.

\subsection{Example\label{sec:mgotraining}}

As a model system to illustrate the performance of CENT, we train an
interatomic potential for magnesium oxide (MgO) 
in the present study. 
MgO is a strongly ionic material which crystallizes in the rock-salt structure
with a wide band gap of about $7.8$~eV,
and is predominantly used as a refractory material due to its high thermal stability.
Our ANN potential for MgO is then used to 
demonstrate some of the key capabilities of FLAME throughout this manuscript.

\begin{figure}[htb]
\begin{lstlisting}[frame=single]
main:
    task:         ann
    types:        Mg O
ann:
    subtask:      train
    approach:     cent1
    optimizer:    rivals
    nstep_opt:    15
    nconf_rmse:   400
    ampl_rand:    0.02
    symfunc:      only_calculate
potential:
    potential:    ann
    ewald:  
        ewald:    False
\end{lstlisting}
\caption{A sample input \texttt{flame\_in.yaml} file to train an ANN model within the CENT scheme. The \texttt{main} block determines the \texttt{task} (\texttt{ann}) and the chemical system (here, Mg--O). The \texttt{ann} block determines the  \texttt{subtask} to be performed, namely \texttt{train}ing, together with a range of parameters: 
the \texttt{cent1} scheme is specified for the ANN \texttt{approach} using the Kalman \texttt{optimizer} as implemented by \texttt{rivals}~\cite{Kalman} with $\texttt{nstep\_opt}=15$ optimization steps (epochs), while $\texttt{nconf\_rmse} = 400$ structures are randomly selected from the reference set as training data. The keyword \texttt{ampl\_rand} sets the amplitude of the random ANN weights during initialization, and the symmetry functions (\texttt{symfunc}) are computed on the fly (\texttt{only\_calculate}) without reading or writing them to disk. The details of the interatomic potential is given in the \texttt{potential} block. A description of all keywords and their associated parameters can be found in the manual.~\label{fig:codeexample}}
\end{figure}

We generate the reference data set as described in
Sec.~\ref{sec:training} and include 
both clusters and periodic bulk structures.
In addition to stoichiometric MgO we 
also incorporate a significant fraction of non-stoichiometric
compositions.
Hence, the resulting potential is well suited to model neutral 
as well as  charged systems, including clusters, nano structures, and crystalline solids.
To train the ANN weight parameters in the CENT potential 
we split the reference data set 
into a training part and a validation part.
By carefully monitoring the root mean square error (RMSE) 
of the CENT predictions for the validation data set 
we can benchmark the performance and
easily detect issues arising from over-fitting.
A sample input \texttt{flame\_in.yaml} file 
for the CENT fitting process is shown
in Fig.~\ref{fig:codeexample}, and
Tab.~\ref{tab:datasets}
contains the detailed makeup of the reference data set.

\begin{table*}[t]
\centering
\caption{Detailed description of the DFT reference data.\label{tab:datasets}}
\begin{tabular}{c c ccc c ccc c}
\hline\hline
                   & &\multicolumn{3}{c}{Training data set}&&\multicolumn{3}{c}{Validation data set} & \\ \cline{3-5}\cline{7-9}
Composition        & & Cluster & Bulk &  Total & & Cluster & Bulk & Total & \\ \cline{1-9}
Stoichiometric     & &   8,518  &  1,192   &  9,710  & &  1,787   &    248   & 2,035 & \\
Non-stoichiometric & &   8,521  &  5,949   & 14,470  & &  1,863   &  1,350   & 3,213 & \\
Total              & &  17,039  &  7,141   & 24,180  & &  3,650   &  1,598   & 5,248 & \\ \hline\hline
\end{tabular}
\end{table*}

\begin{figure}[htb]
\centering
\includegraphics[width=1\columnwidth]{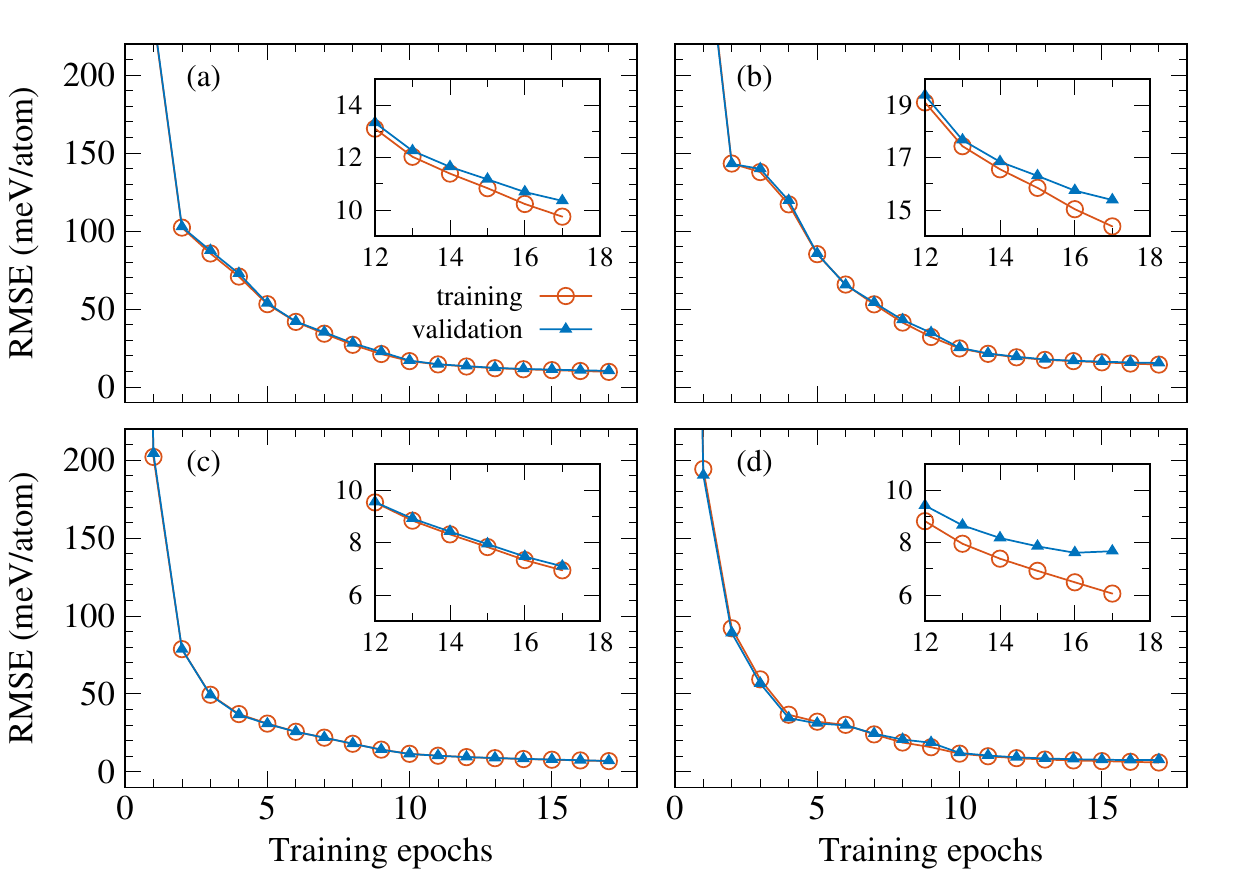}
\caption{The convergence of the RMSE values with respect to
the number of training epochs for (a) all, (b) bulk, (c) cluster,
and (d) stoichiometric (bulk and cluster) structures. Note that
difference in the RMSE between the training and the
validation data is small, in the range of 
1~meV/atom.
\label{fig:rmse}}
\end{figure}

Fig.~\ref{fig:rmse}(a) shows the convergence of the RMSE
with respect to the number of training epochs for all structures in 
the training and validation data sets.
The subfigures~\ref{fig:rmse}(b)-(d) compare the convergence
across structures with different types of BC
and compositions. Overall, the behavior only depends
weakly on the selected subset of training data.
The largest RMSE is observed when considering only
bulk structures, which is however not surprising give that
we included much fewer bulk configurations compared to clusters.

\begin{figure}[htb]
\centering
\includegraphics[width=1.\columnwidth]{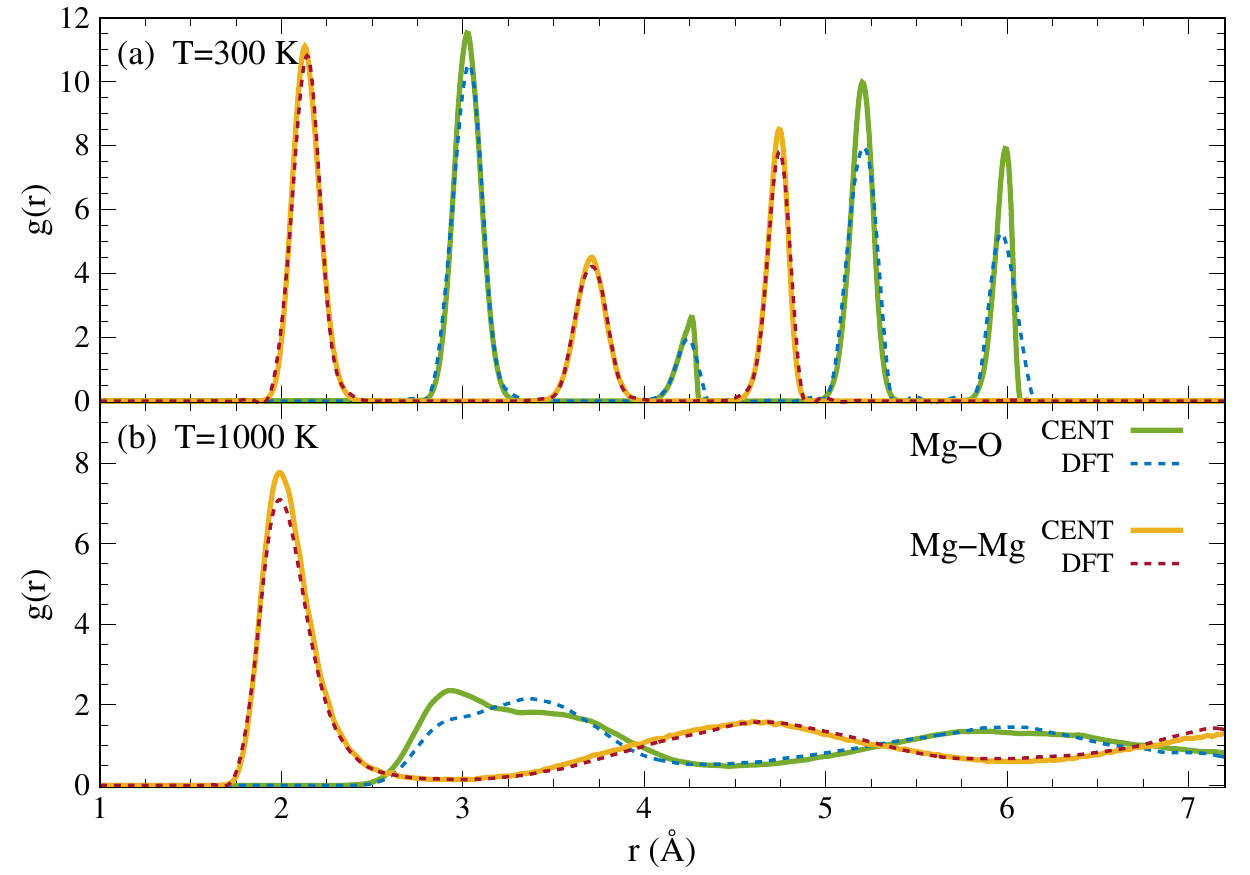}
\caption{The radial distribution functions of the of Mg--O and Mg--Mg interactions computed with
CENT and DFT at finite temperatures of (a) $T=300$~K and (b) $T=1000$~K.~\label{fig:RDF}}
\end{figure}

In order to evaluate the accuracy of the CENT MgO potential
we study the 
interatomic distances at finite temperatures
by performing MD simulations both with CENT
and DFT for a supercell containing $64$ atoms.
We use a Nose-Hoover thermostat
to model a canonical ensemble.
Fig.~\ref{fig:RDF} shows the comparison 
of the radial distribution functions using
CENT and DFT as averaged over the MD trajectories 
at temperatures of $300$~K and $1000$~K after sufficient 
equilibration. 
The overall agreement between the CENT and DFT
results is good, especially in the low-temperature 
regime where the atomic excursions from their 
equilibrium positions is rather small.

\section{Exploring Potential Energy Surfaces\label{sec:PES}}
\subsection{Local Geometry Optimization}
Local minima on the PES correspond to dynamically stable  configurations and define the structural geometry of stable and metastable molecules and crystalline polymorphs. 
Various algorithms to identify such minima in the vicinity of a starting configuration are implemented in FLAME,
ranging from the simple gradient descent approach to quasi-Newton methods and
damped dynamics.

The steepest descent (SD) with energy feedback is the most reliable and numerically stable method available in FLAME. In addition to the
common termination conditions based on either the maximum 
number of iterations or a force tolerance, FLAME offers an additional criterion based on the energy saturation.
The latter termination mode is especially suitable when using SD as a robust precursor to subsequent optimizers that are more efficient but only perform reliably within a quadratic region of the PES.

The conjugate gradient (CG) method is implemented
in conjunction with a line search based on a
quadratic approximation. Although the CG converges 
faster than SD in general, each optimization step 
requires two force evaluations. Two flavors of quasi-Newton
methods are available in FLAME,
namely the stabilized quasi-Newton minimizer
(SQNM)~\cite{Schaefer2015}
and the Broyden-Fletcher-Goldfarb-Shanno
(BFGS) method~\cite{Broyden1970,Fletcher1970} with
different types of (approximate)  line searches. 
Although quasi-Newton methods are very efficient 
they tend to fail if the initial structure is 
not in the vicinity of a local minimum.
The fast inertial relaxation engine~\cite{Bitzek2006} (FIRE)
is an efficient method based on damped dynamics and 
presents a good compromise between reliability 
and efficiency. Its implementation in FLAME works particularly well
for variable cell shape relaxation with 
and without constraints.

\subsection{Structure Prediction\label{sec:structureprediction}}
The task in structure prediction is to identify the lowest energy state on the PES, 
the global minimum, at given conditions. 
Due to the high dimensionality of the PES, 
the search for this ground state presents a
formidable task, especially for large systems with many degrees of freedom.
In particular, the curse of dimensionality 
leads to an exponential increase of the number of local minima
on the PES with respect to the number of atoms in the system~\cite{stillinger_exponential_1999},
which renders the search for the global minimum
extremely challenging. 
In fact, it is impossible to exhaustively map out all
minima for any realistic system,
and sophisticated sampling methods are 
called for to efficiently and thoroughly 
explore the  relevant, low-lying portions of a PES.

\subsubsection{Minima Hopping Method\label{sec:MH}}
The minima hopping method~\cite{Goedecker2004,Amsler2010,amsler_minima_2018} implements a global optimization algorithm
which has proven to be particularly robust and
reliable.
In FLAME, the keyword \texttt{task: minhocao} (Minima Hopping for Crystal Optimization)
in the \texttt{main} block triggers a MHM run.
The MHM employs a sequence of successive short MD runs
and geometry relaxations to ``hop'' between local minima, 
combined with several levels of sophisticated feedback 
mechanisms to learn the features of the PES.
Within each hop, the MHM attempts to escape from the current local minimum, $M_\textrm{cur}$,  using a short MD simulation with a predefined kinetic energy, followed by a local geometry relaxation.
If the escape trial fails, the kinetic energy $E_\textrm{kin}$ is slightly
increase (commonly by around $2-5$\%) to improve the chances to escape
in a the next MD escape step. On the other hand, if the escape trial succeeds, the kinetic energy is slightly reduced. This continuous adaption of $E_\textrm{kin}$ is the first feedback mechanism.

The importance of reducing the kinetic energy upon a successful escape 
lies in the Bell-Evans-Polanyi (BEP) principle~\cite{roy_bell-evans-polanyi_2008}, 
which states that exothermic chemical reactions 
have, on average, low activation barriers. 
Hence, a low kinetic energy during an MD simulation ensures that only low-energy
 barriers can be crossed (corresponding to a low activation energy) due to energy conservation,
 behind which one is more likely to encounter low-energy 
 local minima (corresponding to an exothermic reaction).

A second feedback mechanism controls another energy parameter, called $E_\textrm{diff}$, which introduces an additional preferences for hops towards low-energy structures. A successful hop to  $M_\textrm{new}$ is only accepted if its energy differs from the previous local minimum by less than
a positive threshold value: $E(M_\textrm{new})-E(M_\textrm{cur}) < E_\textrm{diff}$.
The value of $E_\textrm{diff}$ is adjusted in such a way that half of all the performed MHM hops is accepted, while the rest is rejected and a new escape trial is performed, ensuring that even high-energy structures are eventually accepted after sufficiently many steps.

\begin{figure}[htb]
\centering
\includegraphics[width=1\columnwidth]{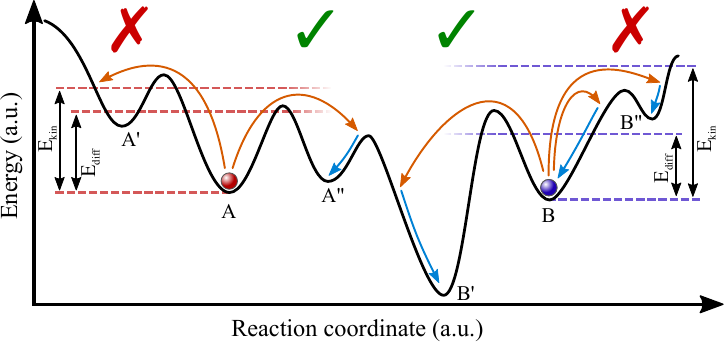}
\caption{A schematic illustration of the working principles of the MHM. The solid black line represents a 1D model energy landscape with its local minima and saddle points. The values of the two feedback quantities $E_\mathrm{kin}$ and $E_\mathrm{diff}$ determine if a hop is possible, and if it is accepted. The hop from A to A' is prohibited due to the lower value of $E_\mathrm{kin}$ compared to the barrier separating the two minima, while the hop to A" is allowed and preferred. The hop from B to B" is rejected since the energy difference of the two local minima exceeds the value of $E_\mathrm{diff}$, while the hop to B' is allowed and leads the system to its ground state.}\label{fig:PES}
\end{figure}

The final feedback mechanism is based on the history of visited local minima. If a known minimum is revisited by the MHM, the kinetic energy is significantly increased. In this way, the system is quickly pushed away from portions of the PES that have already been sampled, towards new, unexplored regions. In fact, the intricate interplay between the parameters $E_\textrm{kin}$  and $E_\textrm{diff}$ ensures that an MHM simulation will never get stuck on any part of the PES: after a funnel has been explored, the kinetic energy will start to increase due to the feedback on $E_\textrm{kin}$. Initially, this will lead to the sampling of high-energy local minima, which will be rejected at first due to the low value of $E_\textrm{diff}$. However, after a while  they will be accepted, and the MHM will leave a
funnel (superbasin) to explore new portions of the PES.
The interplay between the different feedback mechanisms is illustrated in Fig.~\ref{fig:PES}.

The feedback mechanisms above require a reliable method to compare local minima. In FLAME, this comparison is performed based on a combination of energy differences and  structural fingerprints that provide a similarity metric. A detailed description of the available methods in FLAME can be found in Sec.~\ref{sec:descriptors}.  Overall, the collection of feedback mechanisms is an essential part of the MHM which differentiates it from thermodynamics based approaches like Simulated Annealing or Basin Hopping.

\subsubsection{Softening\label{sec:softening}}
Since the MHM hops
between the local minima are essentially
chemical reactions, they correspond to rare events
that occur on a much longer time scale than 
observable through conventional MD simulations.
In fact, the system will frequently merely oscillate 
in the catchment 
basin surrounding a local minimum
during an MD escape trial.
To accelerate the escape from a local minimum,
the MD trajectories must be biased towards a transition
into a neighboring catchment basin.
Hence, we project out the velocity components along
hard modes to essentially ``shoot'' the
system preferably into a direction of low curvature.
This procedure, which is called ``softening'', accelerates
the structural search due to the following reasons.

\begin{itemize}
\item By eliminating the high-frequency vibrations
we can use larger time steps to integrate the
equations of motion for longer time scales,
which directly reduces the computational cost.
\item We can better exploit the BEP: there is a correlation
between the curvature of the eigenmodes in a local minimum and the
height of the transition barrier encountered in that direction.
Hence, starting an MD simulation approximately along a soft
direction increases
the probability of encountering a low barrier,
behind which one is more likely to find a low-energy
structure.~\cite{Sicher2011}
\end{itemize}

FLAME implements a dimer method to identify the low curvature direction at a local minimum.
Initially, the direction of a dimer of length $d$ is chosen by randomly generating
an initial velocity vector based on a Boltzmann distribution
with the current kinetic energy $E_\mathrm{kin}$.
This velocity vector is iteratively rotated by minimizing
the dimer energy with a gradient descent method,
until the curvature along the dimer direction is
sufficiently low or if a predefined amount of softening
iterations is reached. This softening procedure 
is repeated for every MHM step prior to performing the MD escape trial. 
In FLAME, the dimer length $d$ as well as the step size within the dimer 
minimization can be set manually,
or a gradient feedback mechanism
can be used to automatically adjust their values.

The progress of the softening procedure can be monitored in the output of FLAME,
where the dimer energy is reported together with the curvatures
along the dimer direction.
Two approximations of the curvatures are used,
either based on the second order finite difference of the
energies or from the first order finite difference based 
on the force 
acting on the
dimer.
Both approximations will converge to the same value
in the limit of small dimer size $d$ within the harmonic approximation.

The number of softening iterations should be large enough to eliminate
the high-frequency vibration,
but not too large since the dimer will converge to point exactly
along the lowest curvature eigenmode,
thereby sacrificing ergodicity.
As a rule of thumb, we have found that reducing the initial
curvature $\kappa_0$ by one order of magnitude is optimal,
and the softening should be stopped as soon as $\kappa_i<0.1\kappa_0$.

\subsubsection{Example}
\begin{figure}[htb]
\centering
\includegraphics[width=1\columnwidth]{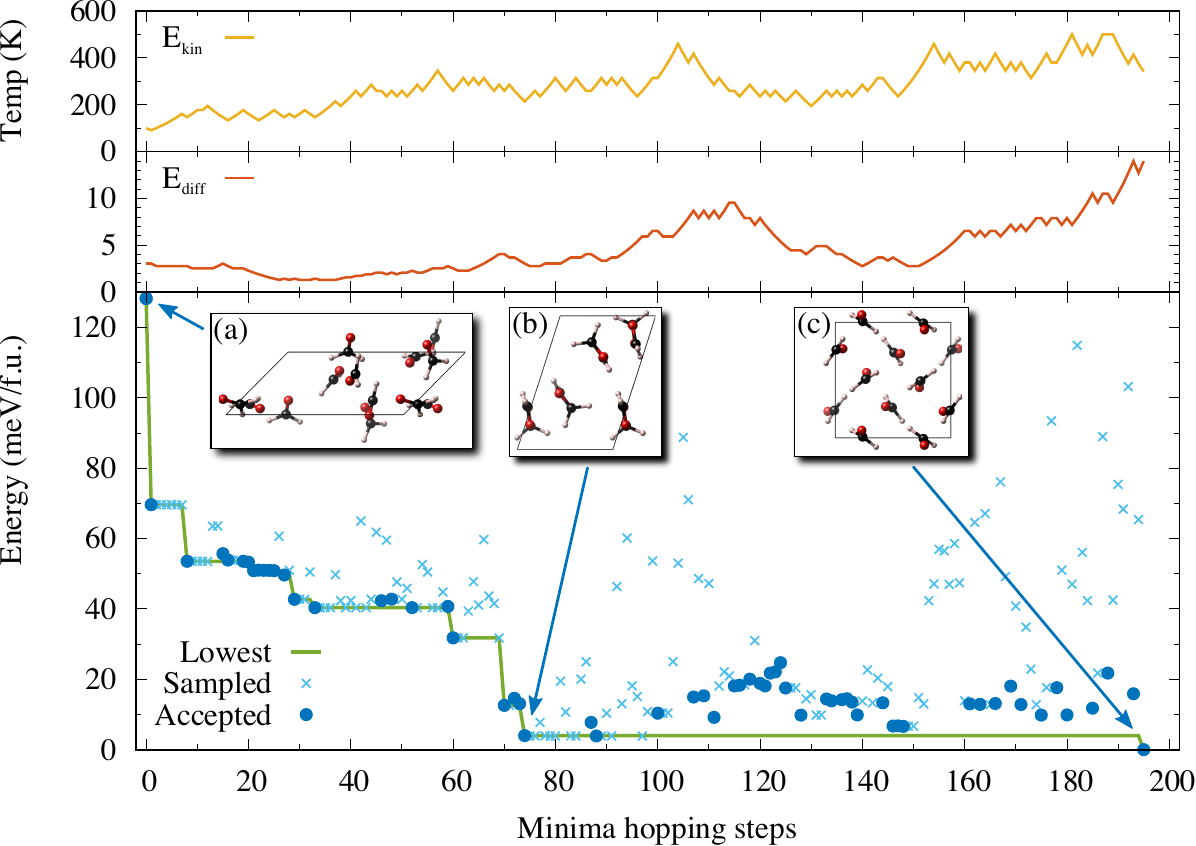}
\caption{The evolution of the parameters during a serial MHM structure prediction run on Formaldehyde. The top and middle panels shows the kinetic energy for each of the MD escape steps and the feedback parameter $E_\mathrm{diff}$, respectively. The bottom panel shows the evolution of the potential energy of the explored local minima. The light blue crosses denote all visited minima, while the dark blue filled circles denote the accepted local minima. The green line represents the lowest energy structure found up to a given MHM iteration.\label{fig:mhmrun}}
\end{figure}

The MHM has been used in the past to predict the structure of a wide range of materials, predominantly of inorganic compounds~\cite{Amsler2009,De2011,Amsler2012,tran_low_2012,Huan2013,SarmientoPerez2015,Amsler2016,ValenciaJaime2016,Asna2017,amsler_prediction_2017,Baledent2018,amsler_exploring_2018,Amsler2019}. However, the implementation of the MHM in FLAME includes a wide range of BC and constraints that can be tuned to optimize the search for any type of material. E.g., individual components of the simulation cell $\{a,b,c,\alpha,\beta, \gamma \}$  can be constrained if experimental lattice parameters are known, or 2-dimensional confinement potentials can be included to model 2D materials or layered structures~\cite{Asna2017,amsler_cubine_2017,singh_low_2019}. Here, we demonstrate the use of FLAME to predict the structure of a molecular, organic crystal, and show how the various parameters evolve during the course of a search in Fig.~\ref{fig:mhmrun}.

We aim to predict the ground state of Formaldehyde, a simple molecule with the chemical formula \ce{CH2O}. Experimentally, the ground state structure of Formaldehyde and its deuterated version has been resolved at 15~K by Weng~\textit{et al.}~\cite{weng_crystal_1989}. Hence, for this particular case, we know what the solution of our search problem is. The crystal structure of the ground state has $P\bar{4}21c$ symmetry (space group index 114), and its unit cell contains 8 f.u. To model the atomic interaction in FLAME, we employ a semiempirical method, the density functional tight binding method as implemented in the DFTB+ package~\cite{aradi_dftb+_2007}. We include Van der Waals interactions in our PES, modeled by empirically fitted pairwise potentials, and employ a sufficiently dense k-point sampling to converge the total energy.

The initial seed structure for our MHM is generated randomly, and is shown in the bottom panel of   
Fig.~\ref{fig:mhmrun}, inset (a). This structure is quite high in energy with respect to the ground state at above 120~meV/f.u., so we can assume that it is rather far away from the ground state structure also in configurational space. We further initialize the two main feedback parameters, $E_\mathrm{kin}$ and $E_\mathrm{diff}$, with conservative values of 100~K and 2.5~meV/f.u., respectively. Note that we give the kinetic energy here in units of a temperature (an intensive quantity), but it does not correspond to a real temperature since (a) the MD escape trials are very short bursts solely intended to overcome reaction barriers and (b) the relative masses of the atoms are scaled to reduce the range of the frequency spectrum stemming from all vibrational modes.

Since we are dealing with a molecular crystal we have to preserve the molecular units and avoid dissociation of the individual molecules. One way to impose this constraint is to treat each molecule as a rigid object, thereby significantly decreasing the degree of freedom (DOF): every molecule has 6 DOF (3 rotational and 3 translational) if treated as a rigid body, whereas there are $3\times N_\textrm{Nat}=12$ DOF if we take into account all internal coordinates in \ce{CH2O}. Here, we choose an alternative method, namely projecting out the strong intramolecular vibrations in the MD escape trials by performing a relatively large, fixed number of softening steps for the atomic degrees of freedom, $\texttt{nsoften}=40$. In this way, the kinetic energy will be naturally distributed to perform molecular or global moves in the configurational space without breaking any intramolecular bonds.

Within the first few dozen MHM steps, the energy of the explored local minima decreases rapidly (blue crosses in the bottom panel of Fig.~\ref{fig:mhmrun}), many of which are accepted (dark blue circles). This is the most favorable behavior of the MHM, and rapid progress is made in exploring new, low-energy local minima. During this part of the MHM,  the parameters  $E_\mathrm{kin}$ and $E_\mathrm{diff}$ do not change significantly, simply fluctuating slightly around some equilibrium value, as shown in the top two panels of Fig.~\ref{fig:mhmrun}. This behavior changes after reaching step 75 in the MHM, when a very favorable structure with nearly ideal molecular packing is found (inset (b) in Fig.~\ref{fig:mhmrun}). For the next $\approx 100$ steps, no lower energy minimum is found, and the simulation is stuck in a large funnel of the PES. As expected, the feedback on $E_\mathrm{kin}$ and $E_\mathrm{diff}$ kicks in, and their values start to increase, thereby driving the system towards new, unexplored regions of the PES. Eventually, after visiting 196 structures, the MHM algorithm escapes from the ``wrong'' funnel and successfully finds the ground state, shown in inset (c) of Fig.~\ref{fig:mhmrun}.

\subsection{Transition State Searches}
Transition states or saddle points are stationary points of the
PES at which the Hessian matrix has
all but one positive eigenvalues. 
Saddle points on the PES are
important since they determine the kinetics 
of chemical reactions and phase transformations
according to transition state theory~\cite{Eyring1935}.
In fact, two neighboring local minima on a PES can be connected through
the MEP,
along which a saddle point always denotes the state of highest energy
and determines the reaction rate of this particular transition.
Identifying a saddle point on a PES is overall more challenging than
finding a local minimum,
since there is no associate target function 
that can be optimized by rigorously following
its gradients.

Transition state search methods are classified into two types:
(i) methods that identifies a saddle point close to 
an initial configuration on the PES, thereby neglecting which minima it connects, and
(ii) methods that aim at finding a saddle point 
connecting two known, usually neighboring, local minima, thereby also
attaining the associated MEP.
In this manuscript, we denote types (i) and (ii) as one-sided
and two-sided methods, respectively.
Typically, one-sided methods follow the minimum mode by inverting the gradient component,
in an approximate manner, along the eigenvector associated with the lowest eigenvalue.
The dimer method~\cite{Henkelman1999} is a prominent example of this type,
while the nudged elastic band (NEB) equipped with the
climbing-image approach~\cite{Henkelman2000} represents a two-sided method.
Both above methods are available in FLAME, together with 
two additional techniques that are discussed in detail below.

\subsubsection{Enhanced Splined Saddle Method\label{sec:splsad}}
The main issue that prevents a smooth convergence
towards a saddle point between two minima is the absence of a target function to optimize.
The splined saddle method~\cite{Granot2008} developed by Granot and Baer,
and later improved by Ghasemi and Goedecker~\cite{Ghasemi2011}, 
alleviates this issue by mapping the problem 
onto convex optimization task.
It splits the saddle point search into two nested optimization
tasks where the inner loop is a one-dimensional global maximization
along the pathway that is embedded in the outer loop which minimizes 
the maximum point obtained in the inner loop.
The method utilizes splines to characterize the pathway,
hence the energy of the maximum point (EMP) along the path
is a function of the coordinates of anchor points.
The gradient of the EMP with respect to the anchor points
can thus be calculated using the equations given in Ref.~\cite{Ghasemi2011}
and its appendix.
With the gradient of the target function at hand,
any gradient-based optimizer can be directly employed
to find its extremum.
Critical to this method is the confidence in finding
the global maximum along the pathway which in turn
can severely affect the reliability of the target function
and its gradient.
As shown in Ref.~\cite{Ghasemi2011}, the method can
be very efficient and outperform competing algorithms.
However, the method may be unstable for long pathways,
in particular when there exists an intermediate minimum along the reaction trajectory.

\subsubsection{Bar-Saddle Method\label{sec:barsaddle}}
The bar-saddle method~\cite{Schaefer2014} is primarily suited for one-sided
searches and is a modification of the dimer method~\cite{Henkelman1999}.
A bar (or dimer) is moved on the PES 
in such a way that its center converges towards a saddle point
while its orientation points along the direction of 
lowest (negative) curvature. To achieve this, the forces acting on 
its end points are decomposed and used to iteratively rotate and translate the bar, eventually moving its center towards the saddle point.

The main difference between the bar-saddle approach and the dimer
method lies in how the rotational and translational forces are computed. 
Within the dimer method the rotation is expressed in terms of
the rotational angle $d \theta$ within the plane of rotation, given the force
acting perpendicular to the dimer.
In contrast, the bar-saddle approach 
applies a rotational force on the two endpoints of the  bar, $A$ and $B$,
according to  $\textbf{F}_{A}^{\text{Rot}}=\frac{1}{2}(\textbf{F}_{A}^{\perp}-\textbf{F}_{B}^{\perp})$ and $\textbf{F}_{B}^{\text{Rot}}=\frac{1}{2}(\textbf{F}_{B}^{\perp}-\textbf{F}_{A}^{\perp})$,
where $\textbf{F}_{\text{i}}^{\perp}=\textbf{F}_{\text{i}}-\textbf{F}_{\text{i}}^{\parallel}$ are the force
components perpendicular to the bar ends.
For the translational forces, the dimer method 
uses the inverted gradient component
along the dimer, averaged over the two endpoints.
The bar-saddle method on the other hand
uses a cubic interpolation to estimate the force acting along
 the bar, thereby assuming
that the negative mode at the saddle point is harmonic.
After moving the bar using the combination of rotational and translational forces, the bar length is rescaled to its target value, $d_{bar}$. 
 \begin{figure}
 \centering
 \includegraphics[width=1\columnwidth]{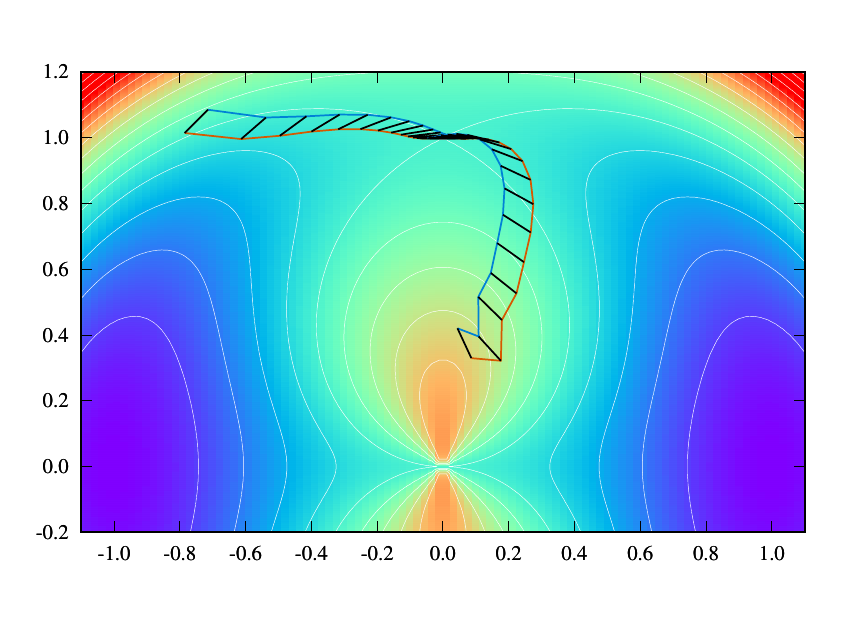}%
 \caption{\label{fig:barsaddle} Trajectory of the barsaddle method starting from two different initial positions on a model energy landscape  $f(x,y)=(1-(x^2+y^2))^2+(y^2)/(x^2+y^2)$. The two local minima (dark purple) are located at $(\pm1,0)$, and the saddle point is located at $(0,1)$.}
 \end{figure}

Since the bar-saddle method is implemented as a one-sided approach in FLAME, the initial 
position of the bar has to be provided as an input and should be ideally in the vicinity of a saddle point.
The dimer is iteratively optimized using a gradient feedback method. Additionally, the 
bar size can be contracted towards the end of a search. 
In this way, the saddle point search starts out with a rather long bar
to approximately locate the saddle point, and hones in 
on its exact location as the gradients on the bar decreases and the
search converges. Fig.~\ref{fig:barsaddle} shows a model PES
together with the iterative steps of the barsaddle method
as it locates the saddle point, starting from two initial 
configurations.

\subsubsection{Example}

\begin{figure}[h]
\centering
\includegraphics[width=0.95\columnwidth]{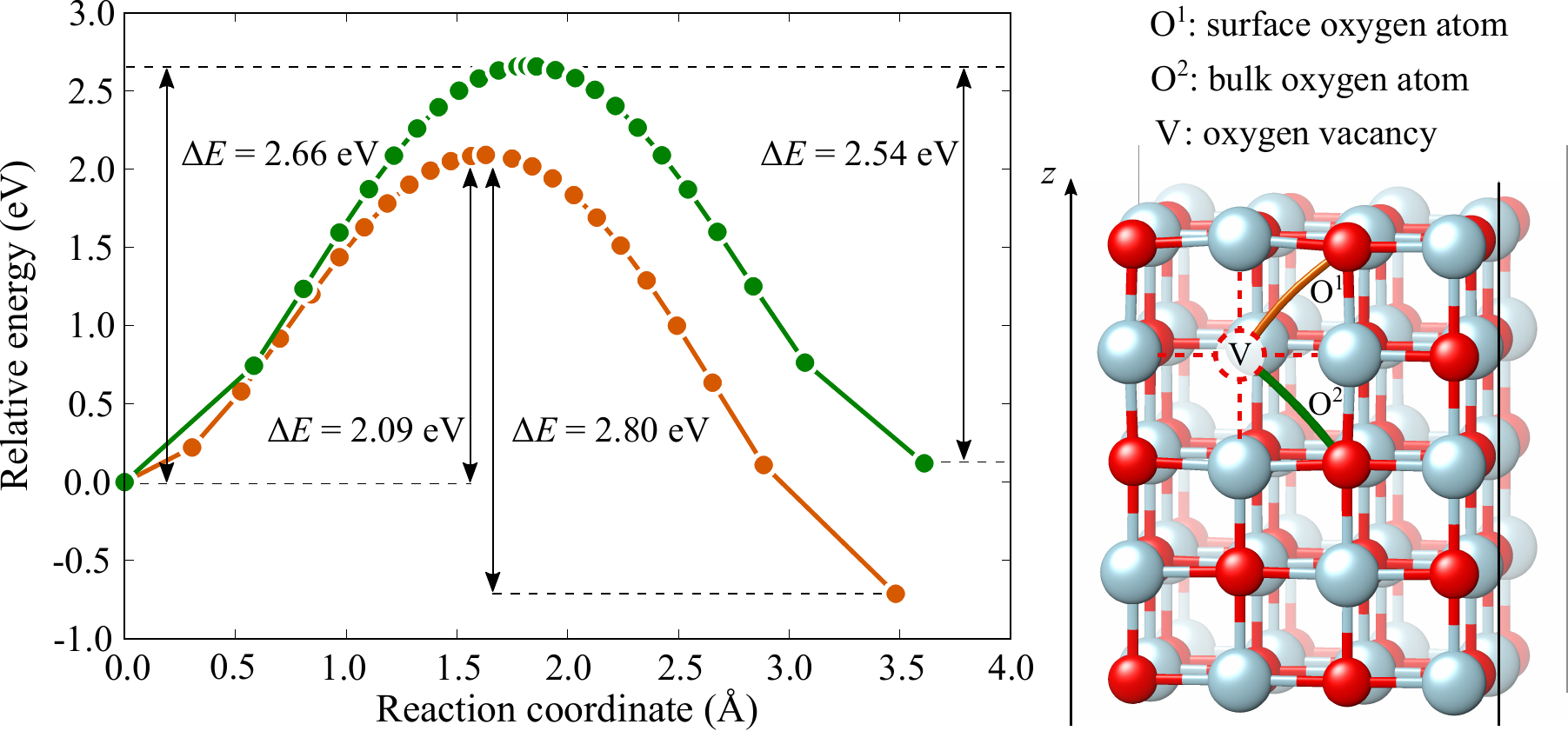}
\caption{Oxygen vacancy diffusion barriers computed with the enhanced 
splined saddle method. The left panel shows the energy along 
the MEP for the
diffusion of the vacancy towards the surface and the center of the bulk (opposite $z$ directions). The right panel shows the model slab structure used in our calculation. The surface normal points along the $z$ direction.}\label{fig:saddle}
\end{figure}

To demonstrate the saddle point optimization
in FLAME, we study the diffusion of an oxygen vacancy near the surface
of a MgO slab.
Two of the diffusion mechanisms, one towards the surface (\ce{O^1})
and the other towards the bulk region (\ce{O^2}), are shown in Fig.~\ref{fig:saddle}.
The oxygen vacancy moving to the surface of the slab passes over a
barrier that is by $\approx 0.7$~eV lower compared to the diffusion towards the bulk.
However, both processes involve crossing high barriers which
may not be surmounted at room temperature in time scales
occurring in typical experiments.
The splined saddle method is used in these calculations
which can be invoked by setting the \texttt{task} key in the  \texttt{main} block to \texttt{saddle} and \texttt{method} key to \texttt{splined\_saddle}
in the \texttt{saddle} block.

\section{Atomic Environment Descriptors and Structural Fingerprints\label{sec:descriptors}}
Structural descriptors of a system or the local environment of its components, the atoms, 
have to be translationally and rotationally invariant.
Further, atomic environment descriptors have to be invariant under 
the permutation of the atoms.
Hence, Cartesian coordinates are not well suited, and alternative schemes are called for
to feed the input layers of an ANN potential,
or to define a metric in configurational space to compare structural differences.

A common approach is to map the Cartesian coordinates into
a so-called constant-sized vectorial environment descriptor,
which fulfills all required symmetry conditions.
Behler introduced a suitable environment descriptor,
the atom-centered symmetry functions (ACSFs)~\cite{Behler2011},
which are constructed by summing up smooth two-body and three-body
functions.
The former gives information on radial distribution
surrounding each atom within a localization region,
while the latter provides also information about the angular arrangements.
Several other descriptors, e.g., the smooth overlap of
atomic positions~\cite{Bartok2013} (SOAP) by
Bart{\'o}k \textit{et al.} or the
overlap matrix (GOM)~\cite{Sadeghi2013,Zhu2016}
by Li \textit{et al.} have been developed, and the latter is available in the FLAME code.

The above descriptors can also be used as structural fingerprints to define a metric $d(a,b)$ in configurational space and to compare two different structures $a$ and $b$. Since the distance between these two structures must be invariant under the exchange of any two atoms in the respective structures, the distance must be minimized over all possible permutations $P$ that matches atom $k$ in $a$ with atom $P(k)$ in b:
\begin{equation}
	d(a,b) = \underset{P}{\min}\sum_k^{N}g(\mathbf{V}_k^a,\mathbf{V}_{P(k)}^b)
\end{equation}
where $\mathbf{V}_i^x$ is the atomic environment vector of atom $i$ in structure $x$, and $g(\mathbf{p},\mathbf{q})$ is a norm defined on the individual atomic environment descriptors $\mathbf{p}$ and $\mathbf{q}$. The ideal permutation is found using the Hungarian algorithm with cubic scaling.

In contrast to such atomic-based methods, global fingerprints integrate out the atomic contributions to give a single descriptive vector for a structure. Such global methods come at a loss of information, but are often faster to compute since the minimization with respect to atomic permutations can be omitted. Structural comparison algorithms can become a bottleneck for global optimization tasks using classical force fields, where thousands of structures are rapidly sampled and have to be efficiently compared. In FLAME, the currently implemented global fingerprints include the Oganov method~\cite{Oganov:2009gr} and the bond characterization matrix (BCM)~\cite{Wang:2012dn,Wang:2015ev}.

\subsection{Example}
 \begin{figure}
 \centering
 \includegraphics[width=1.\columnwidth]{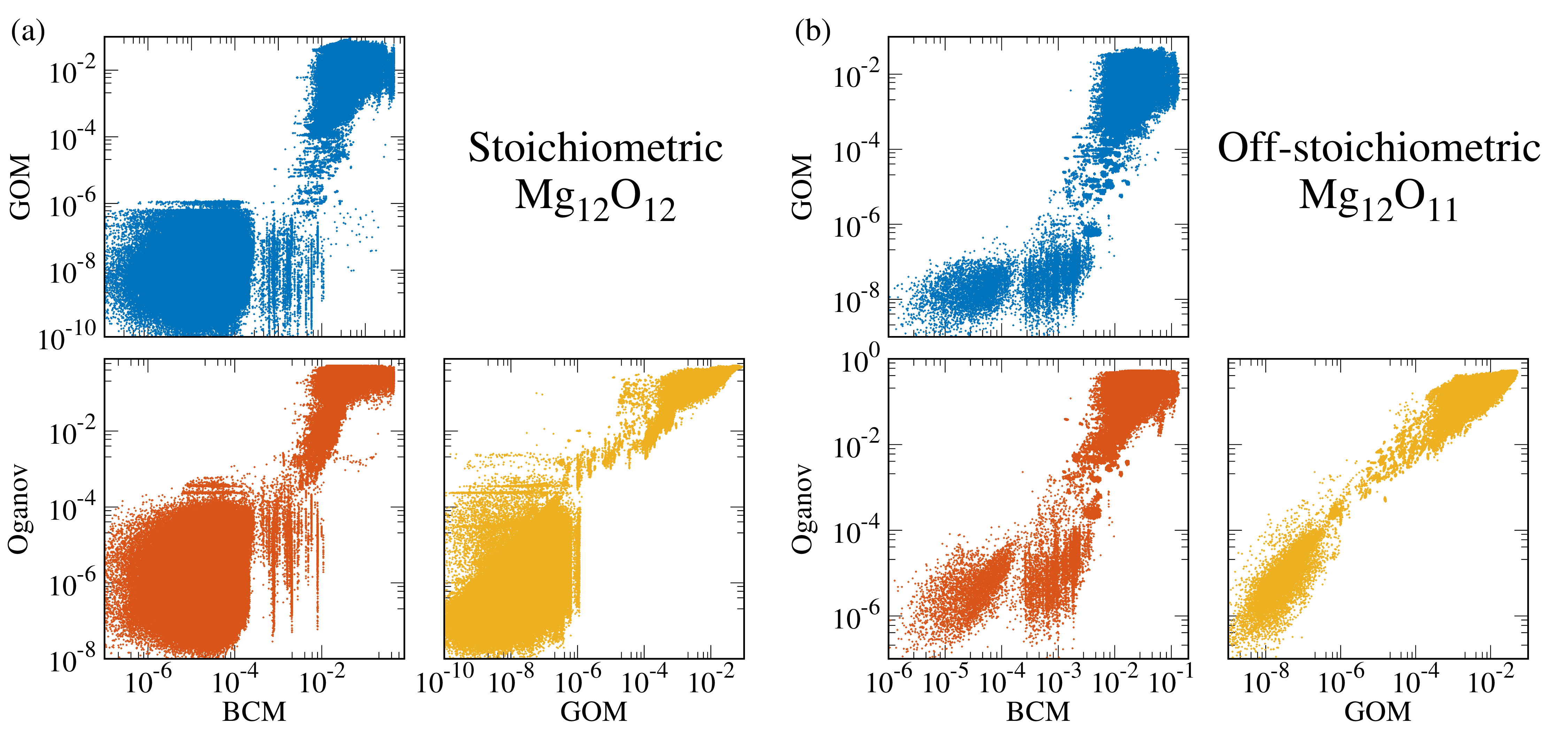}%
 \caption{Pairwise distances using three different structural fingerprints on two periodic Mg--O system. The Oganov and BCM are global fingerprint methods, while the GOM constitutes an atomic-based approach. The units on the axis are arbitrary. \label{fig:FP}}
 \end{figure}

We compare three different environmental descriptor metrics by analyzing their performance on two periodic systems, \ce{Mg12O12} and the off-stoichiometric \ce{Mg12O11}. For both systems we generate a wealth of candidate structures using the MHM and the CENT potential. Using several random input structures, we perform a total of around 1,500 MHM hops for  \ce{Mg12O12} and  \ce{Mg12O11}. Note that we include both the accepted and rejected local minima of the MHM runs as well as all failed escape trials, thereby including many potential duplicate structures. In FLAME, the fingerprints and their distances can be computed on-the-fly during a simulation, or as a post-processing task for data filtering and pruning. Hence, we \textit{a posteriori} compute the fingerprint distances between every pair of structures with respect to the  Oganov~\citep{Oganov:2009gr}, BCM~\cite{Wang:2012dn,Wang:2015ev}, and GOM~\cite{Sadeghi2013,Zhu2016} method.

In Fig~\ref{fig:FP} we show all possible combinations of the pair-wise distances for the two data sets, where different fingerprint methods are plotted along the $x$ and $y$ axis. 
Panels (a) and (b) correspond to \ce{Mg12O12} and  \ce{Mg12O11}, respectively.
Ideally, two fingerprints that perform similarly would correlate well and produce a diagonal line, with a clear gap separating 
structures that are classified as identical and distinct. 
For the \ce{Mg12O12} system we clearly see that all three fingerprints identify a large number of structures that are identical, shown be the large points cloud at the bottom left. This large region corresponds primarily to rock-salt structures and slightly defective versions thereof, indicating that Mg--O is a strong structure seeker with  well-defined global minimum. A close analysis of the subplots shows that there are however some subtle differences between the three methods. E.g., while the GOM--Oganov subplot exhibits a rather clear correlation, both BCM--Oganov and BCM--GOM shows that the BCM has trouble classifying identical structures based on the several sub-clusters, showing up as vertical lines, without a clear gap.

The correlation plots for \ce{Mg12O11} looks slightly different with less pronounced points clouds at the bottom left. Indeed, the PES of the non-stoichiometric \ce{Mg12O11} system is more complex with a less clearly defined ground state. Similar to   \ce{Mg12O12}, the correlation between  GOM and Oganov fingerprint is clearly visible, while the BCM exhibits a rather continuous fingerprint spectrum with ill defined clusters. These results are in good agreement with the findings of Zhu~\textit{et al.}~\cite{Zhu2016}, who performed extensive tests comparing structural difference metrics across various systems. In practice, the GOM or Oganov fingerprint metrics are a good choice for production runs in FLAME, especially due to their much lower computational cost compared to the BCM method.

\section{Electrostatic Interactions\label{sec:electrostatics}}
An important ingredient for the CENT method is 
the calculation of the electrostatic Hartree term.
This can be done either by using one of the methods implemented in FLAME~\cite{Ghasemi2007b,Rostami2016},
or by relying on one of the external solvers implemented in the
\texttt{BigDFT PSolver}~\cite{Genovese2008,Genovese2006,Neelov2007,Genovese2007} library. 
\texttt{BigDFT PSolver} employs appropriate Green's function depending on the BC
to solve the Poisson's equation and therefore avoids introducing any artifact, with a favorable scaling behavior of $\mathcal{O}(N \log(N))$. Depending on the BC, the available solvers in FLAME are as follow.

\begin{description}
\item [Free BC]
The simplest electrostatic method implemented in FLAME is
the pairwise summation according to atomic charge densities.
The method scales $\mathcal{O}(N^2)$ and is only
suited for small systems.
For larger systems we recommend using the 
quasi-linear scaling methods from the 
\texttt{BigDFT PSolver} library~\cite{Genovese2006}.

\item [Slab BC]
Systems with (quasi-)two-dimensional slablike geometries
are typically modeled with periodic BC
in two dimensions and free BC in the third, referred to
as slab BC in this manuscript.
The P$^3$D method~\cite{Ghasemi2007b}, implemented in FLAME,
solves the Poisson's equation while correctly dealing with 
such a slab BC.
Consequently, no vacuum region is required in the
direction perpendicular to the surface,
in contrast to standard plane wave based
Poisson solvers with fully three-dimensional periodicity.
Therefore, highly accurate results can be obtained
even in the presence of non-vanishing dipole moments along the
surface direction.
In particular, the CENT potential in FLAME
can be used to study polar surfaces of materials and their properties.
Also, the P$^3$D method scales as $\mathcal{O}(N\log(N))$
where $N$ is the number of particles in the simulation cell.
In the P$^3$D method, both the charge density and potential
are expanded in plane waves in the two periodic directions, 
while using finite elements in the third dimension. \\

\item [Bulk BC]
The Hartree energy for three-dimensional, fully
periodic systems in the CENT potential can be calculated
either by Fourier summation or by the \texttt{BigDFT PSolver}.
The Fourier summation is calculated using
\begin{multline}
\label{eqn:Ues}
U^\textrm{es} = 
\frac{2\pi}{V}\sum_{\textbf{k}\neq 0}\frac{1}{\textbf{k}^{2}}
\Biggl[
\left(\sum_{i=1}^{N}q_{i}\cos(\textbf{k}\textbf{R}_{i})
\exp(-\frac{\alpha_{i}^{2}\textbf{k}^{2}}{4})\right)^2\\+
\left(\sum_{i=1}^{N}q_{i}\sin(\textbf{k}\textbf{R}_{i})
\exp(-\frac{\alpha_{i}^{2}\textbf{k}^{2}}{4})\right)^2 \Biggr],
\end{multline}
where $V$ is the volume of the simulation cell,
$q_i$ are the atomic charges, and $\alpha_{i}$ are
the widths of the Gaussian atomic charge densities.

The stress tensor in CENT is the sum of the short range part
due to the environment dependent atomic electronegativities, and
the long range part from the Hartree energy.
The former is calculated with
\begin{multline}
\label{eqn:stress_short}
\sigma_{\alpha \beta}^{\textrm{short}} 
= \sum_{i,j=1}^{N} R_{ij}^\alpha F_{ij}^\beta  \\
=-\sum_{k=1}^{N} \sum_{l=1}^{N_{k}} q_{k} \frac{\partial \chi_{k}}{\partial G^{k}_{l}} 
\sum_{i,j=1}^{N} R_{ij}^\alpha \frac{\partial G^{k}_{l}}{\partial R_{j}^\beta}.
\end{multline}
where $R_{ij}^\alpha$ is the component $\alpha$ of $\textbf{R}_{ij}=\textbf{R}_j-\textbf{R}_i$,
and $F_{ij}^\beta$ is the component $\beta$ of the force applied on atom $i$ by atom $j$.
$N_k$ is the number of symmetry functions, indeed the number of nodes in the ANN input layer.
$G^{k}_{l}$ is $l$-th element of symmetry function array of atom $k$.
The latter contribution to the stress tensor is calculated using the derivatives of
the Hartree energy,
\begin{align}\label{eqn:stress_long}
\sigma^\textrm{es} = -\frac{1}{V} \frac{\partial U^\textrm{es}}{\partial h} h^{T}. 
\end{align}
\end{description}

\subsection{Example}

\begin{figure}[htb]
\centering
\includegraphics[width=0.95\columnwidth]{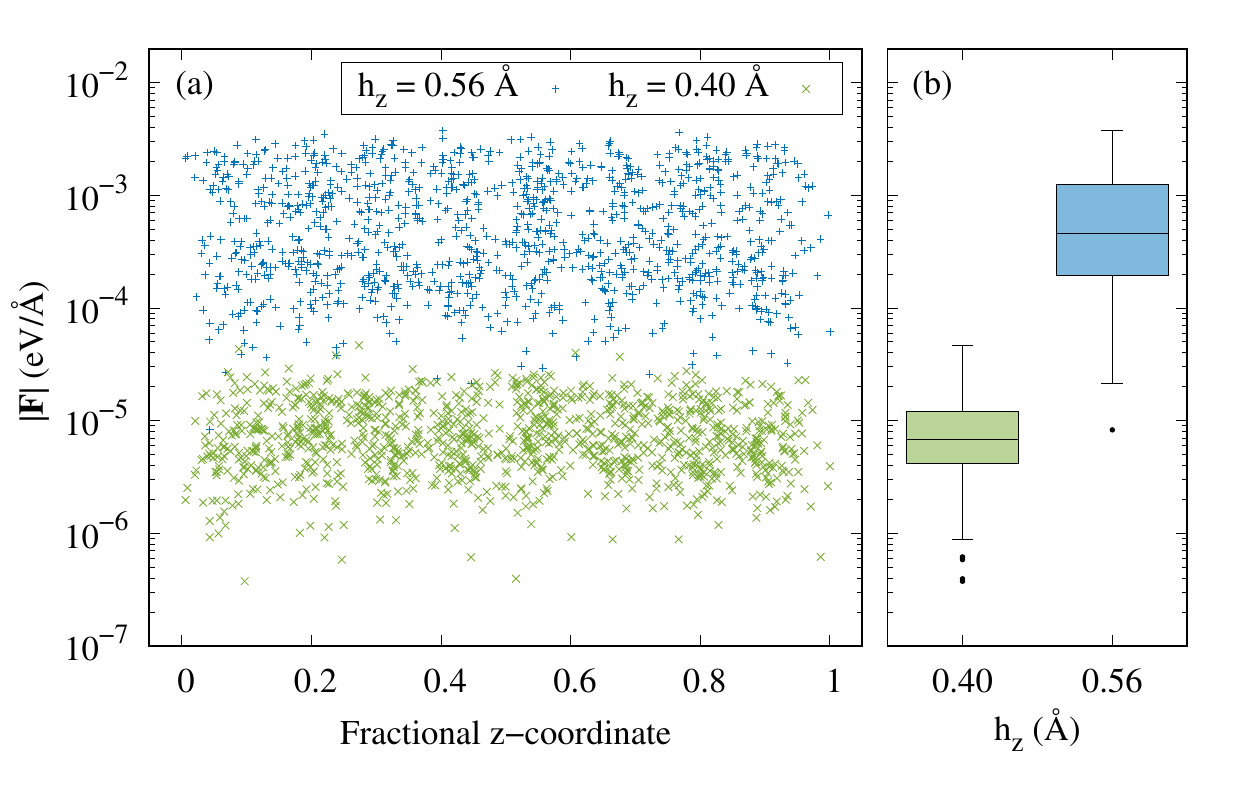}
\caption{Norms of the force errors for each of the atoms with the CENT method within a slab geometry
calculated by the P$^3$D method.
Due to the proper treatment of BC, there is no sign
dependency in the force errors along the direction orthogonal to the surface
of the slab.
Also, it is shown that a small decrease in grid spacing can reduce
the error by more than two orders of magnitude.}\label{fig:p3d}
\end{figure}

To demonstrate the accuracy of the P$^3$D method in evaluating
the electrostatic interaction, we present here an
analysis of its errors when used in 
conjunction with the CENT potential.
More precisely, we investigate how the accuracy of the atomic
forces behaves as a function of the grid density in the 
P$^3$D scheme for a slab of MgO.
To recall, in the CENT potential the charge density is given
by a superposition of atom-centered smooth Gaussian charges.

In order to gather statistically significant results we
generate a set of slab structures by performing an MD
simulation at $1000$~K, starting with a rocksalt-like 
structure consisting of $100$ atoms.
We then randomly select $12$ distinct MD snapshots,
which serve as our benchmark structures.
To generate the reference data we compute the forces 
acting on each atom using fine grid spacings
of  $h_x=h_y=0.29$~\AA\, in the two periodic dimensions,
and $h_z=0.26$~\AA\, along the non-periodic, out-of-plane direction.
With these tight settings the forces are 
converged to essentially within machine precision.

To assess the accuracy of the P$^3$D method we then recompute 
the atomic forces using larger values of $h_z$ while keeping $h_x=h_y=0.29$~\AA\,  fixed.
For most CENT calculations, a spacing of 
$h_z=0.56$~\AA \, is sufficiently small to give 
reliable results of standard accuracy. 
The blue crosses in Fig.~\ref{fig:p3d} show 
the error in the force norms
on each atom as a function
of the $z$-components for all $12 \times 100$ atoms.
Here, the units of $z$ is given in fractional coordinates
with respect to $z_\textrm{min}$ and $z_\textrm{max}$
of the two outmost atoms at the top and bottom of the slabs, respectively.
Note that there is no systematic pattern in the
error distribution along the $z$ direction,
indicating that the error in atomic forces at the surfaces
and at the center of the slab are 
virtually identical.
Such a behavior is crucially important when 
dealing with surfaces and
interfaces, and our results demonstrate that the P$^3$D 
method is particularly well suited to handle these systems.

If a higher accuracy is required, a slight decrease in the 
grid spacing is sufficient: By reducing the value of
$h_z$ to $0.40$~\AA, the errors in the force norms drops by
two orders of magnitude at only a moderate increase in computational cost, as shown by the green crosses in
Fig.~\ref{fig:p3d}. However, for most practical applications 
there is no need to go beyond this level of accuracy.

\section{Conclusions\label{sec:conclusions}}
The use of ML techniques for atomistic simulations
is becoming increasingly popular, and their deployment 
in interatomic potentials can significantly accelerate
and improve theoretical predictions. 
The FLAME code implements the CENT ANN potential
together with a plethora of 
state-of-the art atomistic modeling techniques
in a fully integrated open-source software package.
When compiled as a library, the CENT potential
can be used as a black-box engine and used 
by third-party software packages, like LAMMPS.
On the other hand, the sampling algorithms within FLAME
can readily be linked with external (quantum) engines, 
like LAMMPS, VASP, ABINIT, Quantum ESPRESSO, and many more.
In fact, FLAME can act as a server and communicate
over sockets with any package that supports the i-Pi 
protocol, which has been meanwhile integrated in a range
of codes.

The seamless integration of the CENT potential
with the MHM for structure prediction is a 
particularly powerful feature of FLAME.
The unique combination of a rapid PES exploration
scheme with
an efficient and accurate interatomic potential
has proven to be especially valuable  
in materials discovery.

\section{Acknowledgements\label{sec:ackn}}
We thank Luigi Genovese, Thomas Lenosky, and Stefan Goedecker for fruitful discussions. M.A. acknowledges support from the Novartis Universit\"{a}t Basel Excellence Scholarship for Life Sciences and the Swiss National Science Foundation (projects P300P2-158407, P300P2-174475, and P4P4P2-180669).

\providecommand{\noopsort}[1]{}\providecommand{\singleletter}[1]{#1}%

\end{document}